\documentclass[preprint2]{aastex}
\begin{document}

\title{Searching Beyond the Obscuring Dust Between the Cygnus-Aquila Rifts for Cepheid Tracers of the Galaxy's Spiral Arms}
\author{Daniel J. Majaess, David G. Turner, David J. Lane}
\affil{Saint Mary's University, Halifax, Nova Scotia, Canada}
\affil{The Abbey Ridge Observatory, Stillwater Lake, Nova Scotia, Canada}
\email{dmajaess@ap.smu.ca}

\begin{abstract}
A campaign is described, open to participation by interested AAVSO members, of follow-up observations for newly-discovered Cepheid variables in undersampled and obscured regions of the Galaxy.  A primary objective being to use these supergiants to clarify the Galaxy's spiral nature. Preliminary multiband photometric observations are presented for three Cepheids discovered beyond the obscuring dust between the Cygnus \& Aquila Rifts ($40\le \ell \le 50 \degr$), a region reputedly tied to a segment of the Sagittarius-Carina arm which appears to cease unexpectedly. The data confirm the existence of exceptional extinction along the line of sight at upwards of $A_V\simeq6$ magnitudes ($d\simeq2$ kpc, $\ell \simeq47 \degr$), however, the noted paucity of optical spiral tracers in the region does not arise solely from incompleteness owing to extinction.  A hybrid spiral map of the Galaxy comprised of classical Cepheids, young open clusters \& H II regions, and molecular clouds, presents a consistent picture of the Milky Way and confirms that the three Cepheids do not populate the main portion of the Sagittarius-Carina arm, which does not emanate locally from this region. The Sagitarrius-Carina arm, along with other distinct spiral features, are found to deviate from the canonical logarithmic spiral pattern.  Revised parameters are also issued for the Cepheid BY Cas, and it is identified on the spiral map as lying mainly in the foreground to young associations in Cassiopeia.  A Fourier analysis of BY Cas' light-curve implies overtone pulsation, and the Cepheid is probably unassociated with the open cluster NGC 663 since the distances, ages, and radial velocities do not match.
\end{abstract}

\section{Introduction}
Classical Cepheid variables and young open clusters trace the Galaxy's spiral features in a consistent fashion \citep{wa58,bo59,ks63,ta70,op88,ef97,be06,ma09}. Establishing distances for newly discovered classical Cepheids is therefore useful for further studies of the Milky Way's structure.

The {\it All Sky Automated Survey} \citep[ASAS,][]{po00}, the {\it Northern Sky Variability Survey} \citep[NSVS,][]{wo04}, and {\it The Amateur Sky Survey} \citep[TASS,][]{dr06}, have possibly made over 200 detections of new Cepheid variables through their photometric signatures \citep{ak00,wg04,es07,be09}, a sizeable addition to the Galactic sample \citep{ha85,be00,sd09}.  However, some candidates possess only single passband photometry, which is insufficient for establishing either a distance or color excess through existing relations \citep{ta03,ben07,vl07,fo07,lc07,ma08a,tu09c}.  The present study outlines a campaign of multiband photometry from the Abbey Ridge Observatory \citep{la07,ma08b} to establish mean $BV$ magnitudes for newly-detected classical Cepheids in undersampled regions of the Galaxy. The primary focus lies in long-period classical Cepheids between the Cygnus-Aquila Rifts \citep[$40\degr \le \ell \le 50 \degr$,][]{fo83,fo84,fo85,da01,st03,pr08}, the intent being to trace the Sagittarius-Carina arm through the first Galactic quadrant, where it appears to cease unexpectedly \citep[e.g.,][]{ge76}.  Long-period classical Cepheids are young and massive \citep[e.g.,][]{tu96}, ideal characteristics for objects used to delineate spiral structure, since they have not had time to travel far from their birthplaces in the arms. 

This study also highlights ways in which small telescopes, like those used by most AAVSO members, can contribute to our knowledge of Galactic structure via classical Cepheids. The examples may provide inspiration for enthusiasts to add newly-discovered Cepheid variables to their own observing programs.  On the order of a thousand variables are flagged as suspected Cepheids in the ASAS alone, of which a small yet relevant fraction are \textit{bona fide} Cepheids.  Multiband Johnson mean magnitudes for such objects would enable the determination of distances and reddenings through existing relationships \citep[e.g.,][]{ma08a,tu09c}. Establishing reliable photometric parameters for new Cepheids increases the size of the Galactic sample, thereby helping to place stronger constraints on a host of Galactic parameters, including its warp and spiral structure \citep[e.g.,][]{ma09}.

\begin{figure}
\includegraphics[width=7cm]{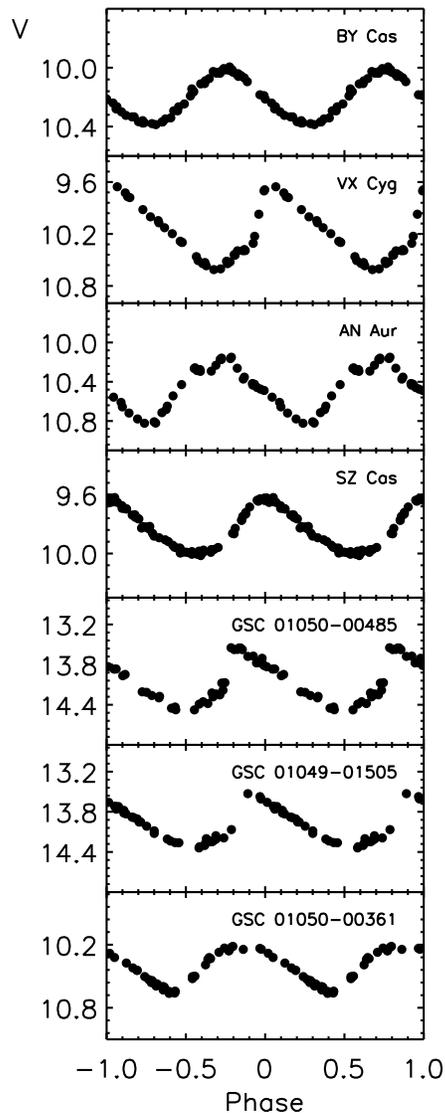}
\caption{\small{Light curves for several Cepheids being monitored from the Abbey Ridge Observatory.}} 
\label{fig9}
\end{figure}

\begin{deluxetable}{lcccccccc}
\tabletypesize{\small}
\tablecaption{Cepheids observed from the Abbey Ridge Observatory.} 
\label{arophot}
\tablewidth{0pt}
\tablehead{\colhead{Star ID} &\colhead{$\ell (\degr)$} &\colhead{P (days)} &\colhead{$V$} &\colhead{$B-V$} &\colhead{$V_a$} &\colhead{$B_a$}  &\colhead{d (kpc)} &\colhead{$E_{B-V}$}}
\startdata
GSC 01050-00485 &$47.61$ &$18.25\pm0.03$ &14.02 &3.02 &0.9 &1.5 &2.0 & 1.9 \\
GSC 01050-00361 &$47.56$ &$8.63\pm0.01$ &10.44 &1.96 &0.42 &0.62 &1.1 & 1.1 \\
GSC 01049-01505 &$45.49$ &$20.84\pm0.03$ & 13.92 & 2.53: & 0.7 & 1.0: & 5.2: & 1.4: \\
BY Cas &$129.55$ &$3.2215\pm0.0006$ &10.38 & --- & 1.29 & -- &1.7 & 0.6 \\
\enddata
\end{deluxetable}
Precise differential CCD photometry for short-period Cepheids is needed for Fourier analysis of their light curves in order to help constrain their pulsation mode \citep{be95,we95,bs98,za05}, an important characteristic affecting distance estimates ($\simeq 30\%$).  The Cepheid BY Cas is provided as a pertinent example of how small telescope photometry can help.  Detailed photometry for the star was obtained to assess its pulsation mode and to examine its possible membership in the open cluster NGC 663, an analysis prompted in part by the study of \citet{us01}. Cluster membership is important for the calibration of classical Cepheid distances, reddenings, period-mass relations, and period-age relations \citep{tu96,tu02,lc07,fo07,ma08a,tu09c}. Classical Cepheids like BY Cas are also useful for establishing the Sun's distance above the plane, as well as to deduce the classical Cepheid scale height \citep{fe68,ma09}.  BY Cas is also identified on a hybrid spiral map of the Milky Way, which was constructed to assess the locations of the Cepheids surveyed within the broader context of the Galaxy.

\section{Cepheid Research from the ARO}
\label{saro}

The Abbey Ridge Observatory \citep{la07,ma08b} is engaged in a campaign aimed at studying Cepheid variables \citep{tu08,tu09b}. Light curves for several Cepheids being monitored are presented in Fig.~\ref{fig9} as an example of what can be achieved by means of small telescopes (e.g., the ARO).   Research consists of monitoring northern hemisphere Cepheids to determine rates of period change, a parameter important for constraining rate of stellar evolution and location within the Cepheid instability strip \citep{tu06}.  Suspected connections between classical Cepheids and open clusters are also being investigated \citep{tu02,ma08a,tu08}.  The objective being to establish additional calibrators for distance, reddening, period-mass, and period-age relations \citep{tu96,tu02,lc07,ma08a,tu09c}. Precise photometry ($\sigma\simeq0.01$ mag.) is being obtained for short-period classical Cepheids to discriminate between fundamental mode and overtone pulsators, especially in previously ambiguous cases \citep[e.g., BD Cas,][]{ma08a}.   A future campaign will aim to establish {\it VI} photometry for Type II Cepheids in globular clusters \citep{cl01,pr03,ho05,ran07,rab07,co08}, the primary objective being to test the metallicity dependence in Cepheid distance relations \citep{ma09}.  

\begin{figure*}
\begin{center}
\includegraphics[width=14cm]{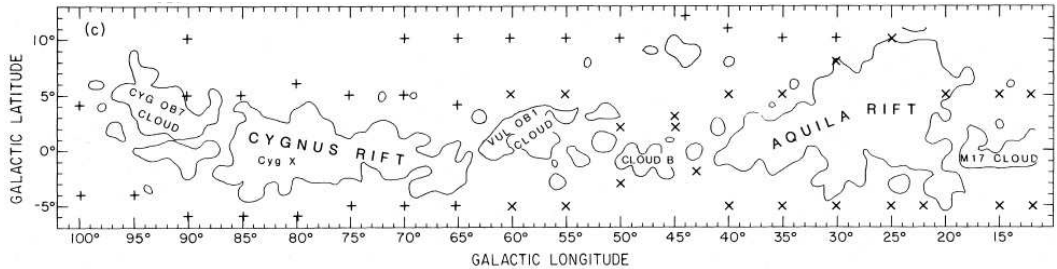}
\caption{\small{Velocity-integrated CO emission map from the seminal work of \citet{dt85}. Much of the region is obscured by molecular clouds, hampering efforts to infer the Galaxy's spiral structure using optical tracers.}}
\label{fig1}
\end{center}
\end{figure*}

The present study highlights our goal to establish mean {\it BV} observations for long-period classical Cepheids discovered in the obscured and undersampled region between the Cygnus and Aquila Rifts (Fig. \ref{fig1}). 

\section{Observations}
\label{sobservations}

All-sky {\it BV} photometry for our program objects were obtained on several nights, with extinction co-efficients derived using techniques outlined by \citet{he98} and \citet{wa06}. The data were standardized to the Johnson system using stars in the open cluster NGC 225 for calibration \citep{ho61}. Period analysis of the photometry was carried out in the \textsc{peranso} software environment \citep{va06} using the algorithms \textsc{anova} \citep{sc96}, \textsc{falc} \citep{ha89}, and \textsc{cleanest} \citep{fo95}. 

\subsection{Cepheids ($40\degr \le \ell \le 50 \degr$)}
\label{lcepheids}

Observations have been obtained for three newly-discovered, suspected, long-period Cepheids from the NSVS \citep[GSC 01050-00485 and GSC 01049-01505,][]{wo04,wg04} and ASAS \citep[GSC 01050-00361,][]{po00}. Phased {\it V}-band light curves for the Cepheids, and BY Cas, are illustrated in Fig. \ref{fig9}, with relevant parameters summarized in Table \ref{arophot}. 

Distances to the Cepheids were computed from a mean of {\it BV} and {\it VJ} reddening-free classical Cepheid distance relations \citep{ma08a}, the latter using mean {\it J}-band magnitudes derived from single epoch 2MASS observations \citep{cu03} following a prescription outlined in \citet{ma08a} \citep[an alternative procedure can be found in][]{so05}. Reddenings for the same stars were estimated from a Cepheid {\it VJ} color excess relation \citep{ma08a}. Single epoch 2MASS infrared {\it J} magnitudes are available for most Cepheids, newly discovered or otherwise. Caution is urged in their use, however, since bright Cepheids ($J\sim5$ mag.) may have saturated infrared magnitudes given their spectral energy distributions. Also, as noted by \citet{ma08a}, the derivation of mean magnitudes from single epoch observations has several complications, one being that Cepheids undergo changes in pulsation period \citep{sz77,sz80,sz81,be94,be97,gl06,tu06}, so a significant time lapse between single epoch observations and those of the reference optical light curves can result in correspondingly large phase offsets. Long-period Cepheids, in particular, tend to exhibit both random and rapid period changes \citep{tb04,tu06,be07,tu09a}.  A large uncertainty in the periods determined for recently discovered Cepheids is a primary concern.  

An additional ten stars from the ASAS are to be monitored as well, with relevant details to appear in a separate study.

\subsection{BY Cas}

The discovery of variability in BY Cas is attributed to \citet{be31} \citep{go94}. A Fourier analysis of the new ARO photometry yields an amplitude ratio $R_{21}=0.069\pm0.011$, a value implying overtone pulsation \citep{we95,bs98,za05}. The distance to BY Cas was computed from a mean of {\it BV} and {\it VI} reddening-free classical Cepheid distance relations \citep{ma08a}, and the reddening was determined using a Cepheid {\it VI} color excess relation formulated from the same study.  Infrared photometry ($I_C$) catalogued by \citet{be00} was used. 

The membership of BY Cas in NGC 663, and thus its inferred distance and pulsation mode, have been questioned previously by \citet{us01}. The implied age \citep[$t\sim10^8$ yrs,][]{tu96} and resulting distance for the Cepheid (Table~\ref{arophot}) are inconsistent with parameters for the cluster NGC 663 \citep[$d\simeq2.8$ kpc, t$\simeq2\times10^7$ yrs,][]{pj94}, a discrepancy also confirmed by recent radial velocity measures of the cluster and Cepheid \citep{li91,go94,me08}. BY Cas is therefore unlikely to be a member of NGC 663, and its parameters obtained through use of the Cepheid relations are preferred (Table \ref{arophot}).  

Lastly, \citet{go94} argue on the basis of radial velocities that BY Cas is a binary Cepheid \citep{sz03}.

\section{The Spiral Nature of the Milky Way}
\label{sspiral}

Maps of the Milky Way's spiral structure exhibit striking differences \citep{ru03,ns03,ben06,ch09,ho09,ma09}. Some 150 years after \citet{al52} first suggested that the Milky Way was a spiral, there is currently no consensus on the number of arms in our Galaxy \citep[see Table 2 of][]{va05}. Moreover, the Sagittarius-Carina and Local arm are not well matched by superposed logarithmic spirals \citep{fo83,ru03,ma09}. That may not be surprising given that an organized and idealistic grand design structure is not a characteristic shared by a sizeable fraction of the universe's spirals, including perhaps the Milky Way. Readers are encouraged to examine images of spiral galaxies observed by HST or catalogued in photographic atlases \citep{sb88} and note that galaxies commonly exhibit arms that branch, merge, twist unexpectedly, and feature a degree of irregularity or flocculence. Furthermore, the possible scenario of the Sun within a spur / Local arm \citep[e.g.,][]{ru03} indicates that such features are likely not unique, and probably exist elsewhere in the Galaxy.

\begin{figure}
\includegraphics[width=7.7cm]{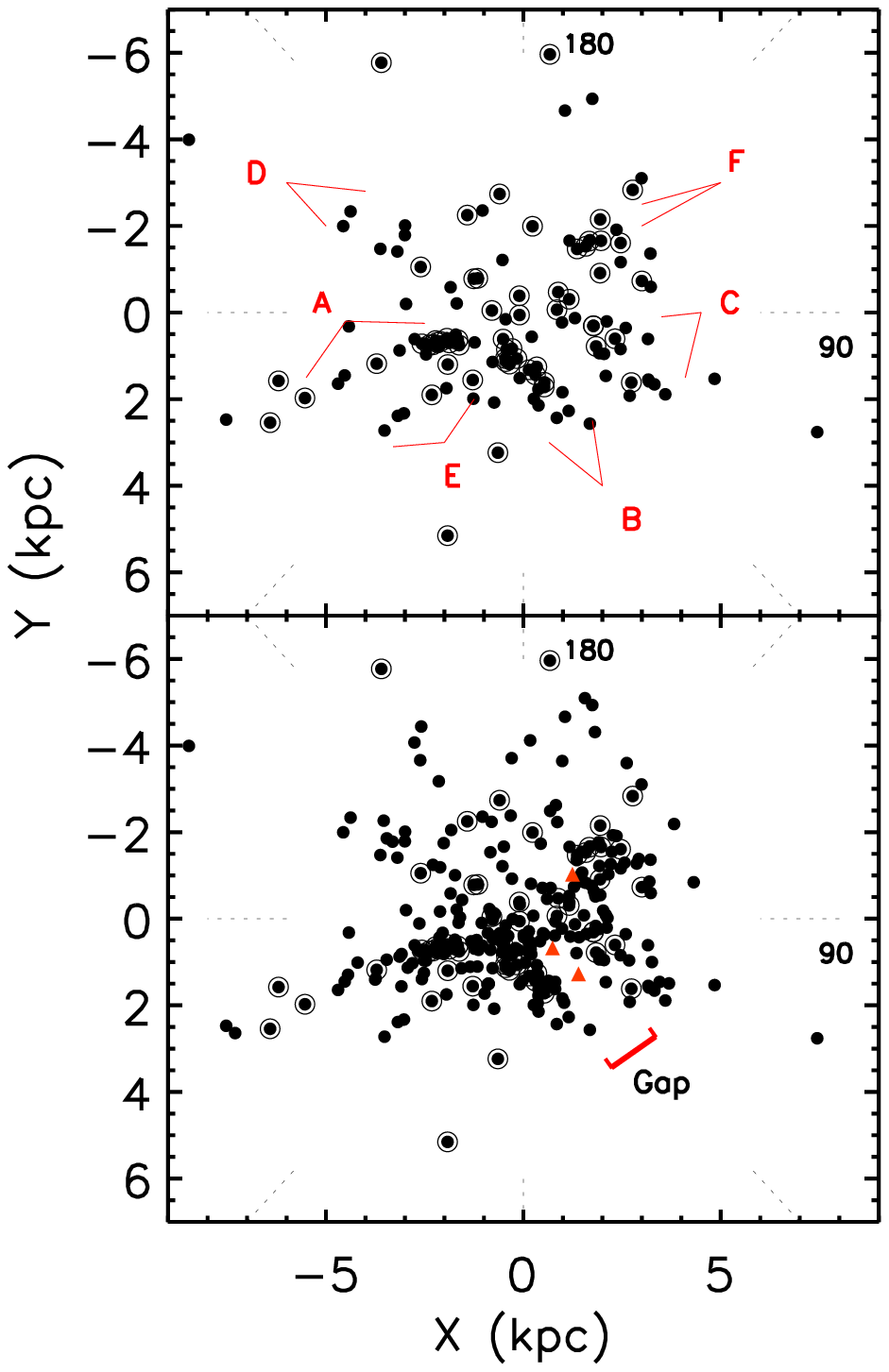}
\includegraphics[width=7.7cm]{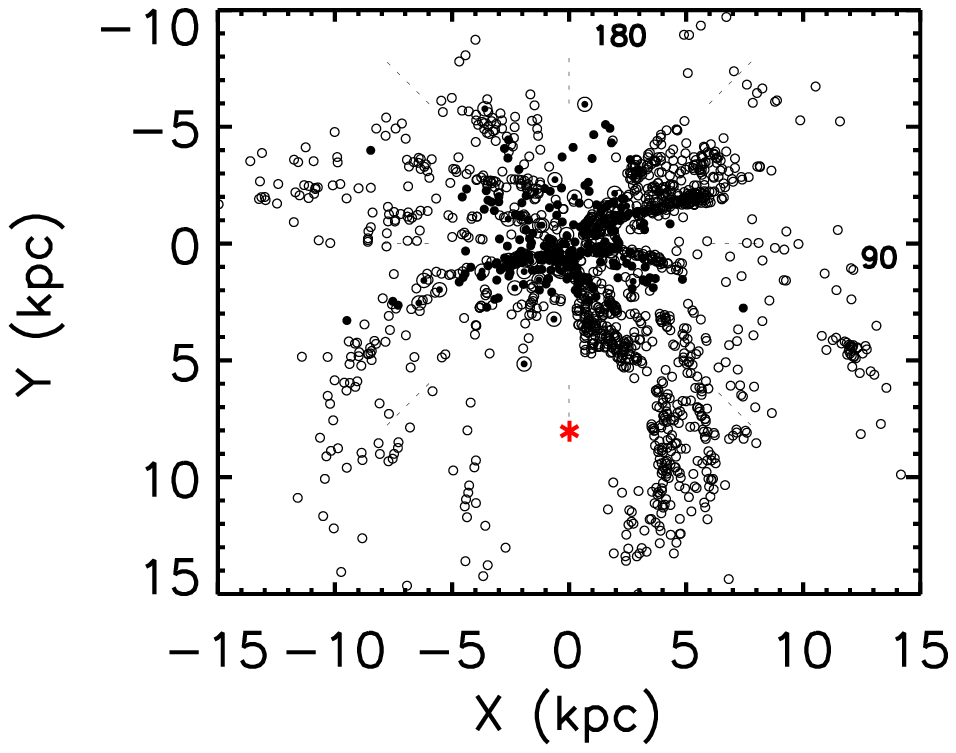}
\caption{\small{Top \& middle, local spiral structure as delineated by classical Cepheid variables (solid points) and young open clusters (YOCs, circled points). See \citet{ma09} for details and the corresponding identifiers. Middle, the gap of optical tracers and locations of the Cepheids studied here are highlighted (\S \ref{sobservations}). Bottom, the Milky Way's structure as illustrated by classical Cepheids, YOCs \& H II regions, and molecular clouds (see text for details). Galactic center is denoted by an asterix \citep{ma09}.}}
\label{fig8}
\end{figure}

\citet{ma09} noted that classical Cepheids \citep[e.g.,][]{be00} and young open clusters \citep{di02,mp03} (YOCs) trace spiral features consistently (Fig. \ref{fig8} top \& middle). The location of the classical Cepheids studied here have been tagged on the classical Cepheid-YOC map (triangles, Fig. \ref{fig8} middle). BY Cas appears to lie foreground to young associations in Cassiopeia (\textit{F}). The new long period Cepheids occupy an obvious gap in the classical Cepheid-YOC data for Galactic longitudes spanning $\ell\simeq35-50 \degr$ \citep{fo83,fo84,fo85,ma09}. The gap has complicated efforts to track the Sagittarius-Carina arm (\textit{A}, Fig. \ref{fig8} top) through this region of the first quadrant, which straddles the Cygnus and Aquila Rifts (Fig. \ref{fig1}).

The Cepheids studied here (\S \ref{lcepheids}) indicate that the extinction is exceptional along this line of sight, partly because of a nearby molecular cloud \citep{fo83,fo84,fo85}. The derived reddeings for the Cepheids at $\ell\simeq47 \degr$ (Table~\ref{arophot}), in tandem with a standard ratio of total to selective extinction \citep[$R = A_V/E_{B-V} = 3.06$,][]{tu76}, implies extinction amounting to upwards of $A_V\simeq3$ magnitudes per kiloparsec, in general agreement with previous results \citep{fo83,fo84,fo85,st03}. A hybrid spiral structure map consisting of classical Cepheids, YOCs and H II regions \citep{ho09}, and molecular clouds \citep{ho09} indicates that the absence of optical spiral tracers in this region is not tied to incompleteness resulting from large extinction (Fig. \ref{fig8} bottom). An alternative delineation of the Sagittarius-Carina arm which deviates from the canonical superposed patterns is again advocated \citep{fo83,ru03,ma09}.  The arm appears to trace along features \textit{A} and \textit{B} (Fig. \ref{fig8}, top) and may continue thereafter along $\ell \simeq 35 \degr$ (Fig. \ref{fig8} bottom). The distinct signature of an additional arm possibly connecting to the Sagittarius-Carina arm can be traced through feature \textit{C} (Fig. \ref{fig8}, top). An abundance of optical tracers near $\ell\simeq0 \degr$ may be a junction that branches into the Carina (\textit{A}) and Centaurus (\textit{E}) features (Fig. \ref{fig8}, top). More work is needed to thoroughly examine the connections. Classical Cepheids, YOCs and H II regions, and molecular clouds otherwise delineate the Milky Way's spiral features consistently at other Galactic longitudes. 

Matching the distribution of classical Cepheids, YOCs and H II regions, and molecular clouds to a standard spiral pattern is rather challenging, especially in consideration of feature \textit{F} (Fig. \ref{fig8}), often treated as a major spiral feature (the reputed Perseus arm). No superposition of such a pattern has been made in Fig.~\ref{fig8}.  Lastly, caution is urged because the spiral map displays features which are somewhat reminiscent of the ``fingers of God effect'' \citep[see ch.12 of][]{sh82}.

\section{Summary}

Described here is a photometric campaign initiated at the Abbey Ridge Observatory aimed at establishing multiband photometry for new Cepheids. The objective is to determine reddenings and distances for key Cepheid variables in undersampled and heavily obscured regions of the Milky Way so to further elucidate any potential spiral structure. A general framework is outlined for AAVSO members interested in conducting similar research.   Preliminary observations and photometric parameters are presented for three classical Cepheids discovered by the NSVS \citep{wo04,wg04} and ASAS \citep{po00} between the Cygnus-Aquila Rifts ($40\le \ell \le 50 \degr$), a region purportedly tied to a segment of the Sagittarius-Carina arm which appears to cease unexpectedly \citep[e.g.,][]{ge76}.  Reddenings inferred from the Cepheids confirm exceptional extinction along the line of sight at upwards of $A_V\simeq6$ magnitudes ($d\simeq2$ kpc, $\ell \simeq47 \degr$), however, it is shown by constructing a hybrid spiral map of the Milky Way from classical Cepheids, YOCs and H II regions, and molecular clouds, that the scarcity of optical spiral tracers in the region is not solely a consequence of incompleteness owing to extinction. Rather, the hybrid map advocates an alternative delineation for the Sagittarius-Carina arm, and many other distinct features, which are not matched by conventional logarithmic spiral patterns.  The three Cepheids surveyed between the Cygnus and Aquila Rifts ($40\le \ell \le 50 \degr$) do not populate the main portion of the Sagitarrius-Carina arm. 

The spiral tracers produce a consistent illustration of the Milky Way (see \S \ref{sspiral}), reaffirming the importance of adopting a multifaceted approach to faciliate an interpretation of the Galaxy's complex structure.  Supplementing optical spiral tracers like classical Cepheids with indicators that are less sensitive to extinction (e.g., molecular clouds) provides larger statistics and more confident conclusions.  

Revised parameters issued for the Cepheid BY Cas place it mainly foreground to the young associations in Cassiopeia (\textit{F}, Fig.~\ref{fig8}).  The Cepheid is argued to be an overtone pulsator based on a Fourier analysis of its light-curve.  BY Cas is probably unassociated with the open cluster NGC 663, which exhibits a different distance, age, and radial velocitiy.

The future release of the ASAS-3N survey shall provide a statistically valid sample of new Cepheids with multiband $VI$ photometry, thereby enabling the distances and reddenings for the variables to be readily determined.  Observatories like the ARO and those run by fellow AAVSO members will continue to serve a role, especially in supplementing the faint end of the survey where the photometric zero-point becomes too uncertain.  The present study supports a tradition of utilizing small telescopes to conduct pertinent Cepheid research \citep{pe80,pe86,sz03b,tu09b}.

\subsection*{ACKNOWLEDGEMENTS}
\small{We are indebted to Leonid Berdnikov \& Laszlo Szabados, whose comprehensive research on Cepheid variables was invaluable to our analysis, to the authors of \citet{ho09} for making the relevant data on H II regions and molecular clouds accessible, Michael Sallman (TASS), Grzegorz Pojmanski (ASAS), Arne Henden and Michael Saladyga (AAVSO), Alison Doane (HCO), Carolyn Stern Grant (ADS), and the folks at CDS. Reviews and books by \citet{el85}, \citet{fm96}, \citet{fe99,fe01}, \citet{fe76,fe02}, \citet{ho02}, and \citet{sz06}, were useful in the preparation of this work.}

\end{document}